\documentstyle[12pt,epsfig]{article}

\newcommand{\be}{\begin{equation}}
\newcommand{\ee}{\end{equation}}

\newcommand{\br}{{\bf r}}
\newcommand{\ba}{{\bf a}}

\newcommand{\vp}{\varphi}

\newcommand{\ra}{\rightarrow}

\newcommand{\gm}{\gamma}
\newcommand{\om}{\omega}
\newcommand{\Om}{\Omega}
\newcommand{\dgr}{\dagger}
\newcommand{\lbd}{\lambda}

\begin{document}

\begin{center}

{\Large {\bf Nonlinear dynamics of ultracold gases in double-well
lattices} \\ [5mm]

V.I. Yukalov$^1$ and E.P. Yukalova$^2$} \\ [3mm]

{\it
$^1$Bogolubov Laboratory of Theoretical Physics, \\
Joint Institute for Nuclear Research, Dubna 141980, Russia \\ [3mm]

$^2$Department of Computational Physics, Laboratory of Information
Technologies,\\
Joint Institute for Nuclear Research, Dubna 141980, Russia}

\end{center}

\begin{abstract}

An ultracold gas is considered, loaded into a lattice, each site of
which is formed by a double-well potential. Initial conditions, after
the loading, correspond to a nonequilibrium state. The nonlinear dynamics
of the system, starting with a nonequilibrium state, is analysed in
the local-field approximation. The importance of taking into account
attenuation, caused by particle collisions, is emphasized. The presence
of this attenuation dramatically influences the system dynamics.

\end{abstract}

\vskip 1cm

{\parindent=0pt
{\bf PACS}: 03.75.Kk; 03.75.Lm; 32.80.Pj; 32.80.Qk

\vskip 1cm

{\bf Keywords}: Ultracold gases; Optical lattices; Quasispin models;
Nonlinear dynamics
}

\newpage

\section{Introduction}

Present-day experiments provide the possibility of creating a large
variety of optical and magnetic lattices with loaded ultracold Bose
and Fermi gases [1--6]. One of the main advantages of such systems is
the potentialities of their almost complete control. The parameters 
of the lattices themselves can be varied in a wide range, as 
well as there exists a great diversity of characteristics of ultracold 
gases, both Bose [7--13] as well as Fermi [14--16]. Such a high level 
of control suggests the feasibility of numerous applications. For many 
of these applications, it is important to have a correct description 
of the system dynamic properties.

Recently, the variety of lattices has been expanded with the
experimental realization of double-well optical lattices, each site
of such lattices being formed by a double-well optical potential [17].
Such more complicated double-well lattices possess a rich variety of
thermodynamic equilibrium states [18]. Dynamics of atoms in these complex
lattices should also be more rich, presenting novel possibilities that
could be employed for various practical purposes, for instance, in 
quantum information processing and quantum computing.

Atomic dynamics inside a separate double-well have been studied earlier
[19--21]. The aim of the present paper is to analyze dynamics not inside
a separate double-well, but in the whole double-well lattice, composed
of many double-well sites.

We consider atoms in a double-well lattice, which in equilibrium would
form the Mott insulating state. But initially atoms are loaded in a
nonequilibrium state. Describing their dynamics, it is important to
invoke realistic approximations, taking into account attenuation caused
by atomic collisions. Resorting to a simple mean-field approximation,
usually employed for treating atomic dynamics inside a separate double
well, would result in perpetual oscillations, never coming to a stationary
state corresponding to equilibrium. Therefore, including attenuation is
of crucial importance.

The plan of the present paper is as follows. First, we derive the
effective Hamiltonian for ultracold gases in an insulating double-well
lattice (Section 2). Then we consider the nonlinear evolution equations,
taking account of particle interactions and attenuation, for the dynamic
variables describing particle tunneling, Josephson current, and atomic
displacements (Section 3). In Section 4, we investigate the existence
and stability of attractors, and present the results of numerical
calculations. Finally, Section 5 is conclusion.

\section{Effective Hamiltonian}

The starting point is the standard energy Hamiltonian in the Heisenberg
field representation,
\be
\label{1}
\hat H = \int \psi^\dgr(\br) H_L(\br) \psi(\br) \; d\br \; +
\; \frac{1}{2}\; \int \psi^\dgr(\br) \psi^\dgr(\br')
\Phi(\br-\br') \psi(\br') \psi(\br) \; d\br d\br' \; ,
\ee
in which $\psi(\br)$ is a field operator, where the time dependence, for
brevity, is omitted, but keeping in mind that $\psi(\br)\equiv\psi(\br,t)$
is the Heisenberg field operator; $\Phi(\br)=\Phi(-\br)$ is an interaction
potential; and
\be
\label{2}
H_L(\br) \equiv -\; \frac{\nabla^2}{2m} \; + \; V_L(\br) \; ,
\ee
where the lattice potential $V_L(\br)$ is assumed to have the
double-well structure in each lattice site marked by a lattice vector
$\ba_i$. For generality, we do not specify a particular form of the
interaction potential $\Phi(\br)$, since in experiments it can have
different forms. It can be a local potential [7--11], or a long-range
dipolar potential, or a combination of a contact and dipolar potentials
[22]. The strength of atomic interactions can also be varied in a very
wide range by using Feschbach resonance techniques [23,24].

The lattice is assumed to be sufficiently deep in order to realize the
Mott insulating state, when the wave functions $\vp_n(\br-\ba_i)$, which
are the solutions of the eigenproblem
\be
\label{3}
H_L(\br-\ba_i) \vp_n(\br-\ba_i) = E_n \vp_n(\br-\ba_i) \; ,
\ee
are well localized around the related lattice sites $\ba_i$. Then the
jumps of atoms between different sites are suppressed, as it should be
for an insulating state. However the tunneling between the wells of a
double-well potential, associated to each lattice site, can be arbitrary
and can be regulated by varying the shape of the double well [25].

Expanding the field operator as
\be
\label{4}
\psi(\br) =\sum_{in} c_{in} \vp_n(\br-\ba_i) \; ,
\ee
we can transform Hamiltonian (1) to the representation of operators
$c_{in}$, whose commutation relations are prescribed by the sort of
atoms, which can be either bosons or fermions. In an insulating state,
the filling factor is to be integer, which we set to one. This implies
the homeopolarity conditions
\be
\label{5}
\sum_n c_{in}^\dgr c_{in} = 1 \; , \qquad c_{im} c_{in} = 0 \; .
\ee
Substituting expansion (4) into Hamiltonian (1) and retaining overlaps
up to second order, gives
\be
\label{6}
\hat H = \sum_{in} E_n c_{in}^\dgr c_{in} \; + \; \frac{1}{2} \;
\sum_{i\neq j} \; \sum_{mn m'n'} V_{ij}^{mnm'n'}
c_{im}^\dgr c_{jn}^\dgr c_{jm'} c_{in'} \; ,
\ee
where
$$
V_{ij}^{mnm'n'}  \equiv \; \Phi^{mnm'n'}_{ijji} \; \pm \;
\Phi^{mnn'm'}_{ijij}
$$
is expressed through the matrix elements of the interaction potential
with respect to the wave functions $\vp_n(\br-\ba_i)$.

The system temperature is assumed to be close to zero, so that, of the
double-well spectrum, only two lowest energy levels may be taken into
account. These levels, $E_1$ and $E_2$, are enumerated with the index
$n=1$ and $n=2$. The wave functions of the lowest levels, corresponding
to a double-well potential, are such that the ground-state function is
symmetric with respect to spatial inversion, while the first excited
state is antisymmetric [21],
\be
\label{7}
\vp_1(-\br) = \vp_1(\br) \; , \qquad \vp_2(-\br) = - \vp_2(\br) \; .
\ee
Both functions $\vp_1(\br)$ and $\vp_2(\br)$ are real.

By introducing the pseudospin operators
$$
S_i^x = \frac{1}{2} \left ( c_{i1}^\dgr c_{i1} - c_{i2}^\dgr c_{i2}
\right ) \; , \qquad
S_i^y = \frac{i}{2} \left ( c_{i1}^\dgr c_{i2} - c_{i2}^\dgr c_{i1}
\right ) \; ,
$$
\be
\label{8}
S_i^z = \frac{1}{2} \left ( c_{i1}^\dgr c_{i2} + c_{i2}^\dgr c_{i1}
\right ) \; ,
\ee
we have
$$
c_{i1}^\dgr c_{i1} = \frac{1}{2} \; + \; S_i^x \; , \qquad
c_{i2}^\dgr c_{i2} = \frac{1}{2} \; - \; S_i^x \; ,
$$
\be
\label{9}
c_{i1}^\dgr c_{i2} = S_i^z \; -i \; S_i^y \; , \qquad
c_{i2}^\dgr c_{i1} = S_i^z \; + i \; S_i^y \; .
\ee
In order to better understand the meaning of the pseudospin operators (8),
we may define the left, $c_{iL}$, and the right, $c_{iR}$, operators by
the relations
\be
\label{10}
c_{i1} = \frac{1}{\sqrt{2}} \left ( c_{iL} + c_{iR} \right ) \; ,
\qquad c_{i2} = \frac{1}{\sqrt{2}} \left ( c_{iL} - c_{iR} \right ) \; .
\ee
Then the operators (8) acquire the form
$$
S_i^x = \frac{1}{2} \left ( c_{iL}^\dgr c_{iR} +
c_{iR}^\dgr c_{iL} \right ) \; , \qquad
S_i^y = - \; \frac{i}{2} \left ( c_{iL}^\dgr c_{iR} -
c_{iR}^\dgr c_{iL} \right ) \; ,
$$
\be
\label{11}
S_i^z = \frac{1}{2} \left ( c_{iL}^\dgr c_{iL} -
c_{iR}^\dgr c_{iR} \right ) \; .
\ee
>From here, we see that $S_i^x$ describes the tunneling intensity
between the left and right wells of a double-well potential located
at the $i$-site. The operator $S_i^y$ characterizes the Josephson
current between the left and right wells. And $S_i^z$ is a displacement
operator showing the imbalance between the wells.

Let us introduce the notation
\be
\label{12}
E_0 \equiv \frac{1}{2} \left ( E_1 + E_2 \right )
\ee
and define the following combinations of the interaction matrix elements
$$
A_{ij} \equiv \frac{1}{4} \left ( V_{ij}^{1111} + V_{ij}^{2222} +
2 V_{ij}^{1221} \right ) \; , \qquad
B_{ij} \equiv \frac{1}{2} \left ( V_{ij}^{1111} + V_{ij}^{2222} -
2 V_{ij}^{1221}\right ) \; ,  
$$
\be
\label{13}
C_{ij} \equiv \frac{1}{2} \left ( V_{ij}^{2222} - V_{ij}^{1111}
\right ) \; , \qquad J_{ij} \equiv -2 V_{ij}^{1122} \; .
\ee
Also, we define the tunneling parameter
\be
\label{14}
\Om \equiv E_2 - E_1 + \sum_{j(\neq i)} C_{ij} \; .
\ee
Then Hamiltonian (6) reduces to the effective pseudospin representation
\be
\label{15}
\hat H = E_0 N + \frac{1}{2} \; \sum_{i\neq j} A_{ij} \; - \;
\Om \; \sum_i S_i^x \; + \; \sum_{i\neq j} B_{ij} S_i^x S_j^x \; -
\; \sum_{i\neq j} J_{ij} S_i^z S_j^z \; ,
\ee
in which $N$ is the number of atoms in the lattice.

\section{Evolution equations}

The Heisenberg equations of motion for operators (8) are
$$
\frac{dS_i^x}{dt} = 2 S_i^y \; \sum_{j(\neq i)} J_{ij} S_j^z \; ,
\qquad
\frac{dS_i^y}{dt} = \Om S_i^z \; - \; 2S_i^x \;
\sum_{j(\neq i)} J_{ij} S_j^z \; - \; 2S_i^z \;
\sum_{j(\neq i)} B_{ij} S_j^x \; ,
$$
\be
\label{16}
\frac{dS_i^z}{dt} = -\Om S_i^y + 2 S_i^y \;
\sum_{j(\neq i)} B_{ij} S_j^x \; .
\ee
Our aim is to study the evolution of the averages
$$
x \equiv \frac{2}{N} \; \sum_i < S_i^x> \; , \qquad
y \equiv \frac{2}{N} \; \sum_i < S_i^y> \; ,
$$
\be
\label{17}
z \equiv \frac{2}{N} \; \sum_i < S_i^z> \; ,
\ee
describing the temporal behavior of the tunneling intensity $x=x(t)$,
Josephson current $y=y(t)$, and the imbalance variable $z=z(t)$.

Considering the statistical averaging in Eqs. (16), we need to
choose a decoupling for the pair correlators of spin operators. The
simplest option, which is usually accepted, would be the mean-field,
or semiclassical, approximation, neglecting all pair correlations. Such
a semiclassical approximation does not take into account quantum effects,
due to particle collisions, which cause the attenuation of dynamical
variables.

To take into account such damping effects, we can employ the
local-field approximation that is used in the theory of magnetic
resonance [26,27] and in the calculations of the dielectric response
functions [28]. The idea of this approximation is the existence of
a kind of local equilibrium even in strongly nonequilibrium systems
[29,30]. Particle collisions are considered as occurring in an
effective local field of other particles [26]. Technically, for the
system we consider here, the local-field approximation is realized as
follows. First, one considers equilibrium mean-field solutions for
Hamiltonian (15). Then, introducing the notation
\be
\label{18}
J \equiv \frac{1}{N} \; \sum_{i\neq j} J_{ij} \; , \qquad
B \equiv \frac{1}{N} \; \sum_{i\neq j} B_{ij} \; ,
\ee
we have
$$
< S_i^x > \; = \; \frac{\Om-2B<S_i^x>}{2D} \; {\rm tanh}
\left ( \frac{D}{2T}\right ) \; ,
$$
\be
\label{19}
< S_i^y> \; = \; 0 \; , \qquad
< S_i^z > \; = \; \frac{J <S_i^z>}{D} \; {\rm tanh}
\left ( \frac{D}{2T}\right ) \; ,
\ee
where $T$ is temperature and
$$
D \equiv \sqrt{(\Om - 2B<S_i^x>)^2 + 4J^2<S_i^z>^2} \; .
$$
The same form of solutions (19) is assumed to hold locally in time
and space, when the system is in local equilibrium. This means that,
in local equilibrium, the average variables (17) are of the form
$$
x_t = \frac{\Om-Bx}{D_t} \; {\rm tanh}\left (
\frac{D_t}{2T} \right ) \; ,
$$
\be
\label{20}
y_t=0 \; , \qquad z_t = \frac{Jz}{D_t}\; {\rm tanh}\left (
\frac{D_t}{2T} \right ) \; ,
\ee
in which
$$
D_t \equiv \sqrt{ (\Om-Bx)^2 + J^2 z^2} \; .
$$
The evolution equations for variables (17) are obtained by averaging
Eqs. (16), decoupling the pair spin correlations, but including the
damping, caused by particle collisions, so that the variables would
attenuate to their local forms (20). This procedure yields the
evolution equations
$$
\frac{dx}{dt} = J yz - \gm_2(x-x_t) \; ,
$$
$$
\frac{dy}{dt} = \Om z - ( J + B ) xz - \gm_2 ( y - y_t ) \; ,
$$
\be
\label{21}
\frac{dz}{dt} = -\Om y + Bxy - \gm_1 ( z - z_t ) \; ,
\ee
where the local fields are given in Eqs. (20). The attenuation
parameters $\gm_1$ and $\gm_2$ can be calculated in the same way
as for spin systems [27], being expressed through the particle
interactions. In our case, these parameters are of order $J+B$.

Before solving Eqs. (21), it is convenient to reduce them to
dimensionless notation. To this end, we define the dimensionless
parameters
\be
\label{22}
\om \equiv \frac{\Om}{J+B} \; , \qquad
b \equiv \frac{B}{J+B}
\ee
and notice that $J/(J+B)=1-b$. We also define the dimensionless
function of time
\be
\label{23}
h \equiv \frac{D_t}{J+B} \; .
\ee
For the dimensionless damping parameter, we take
\be
\label{24}
\gm \equiv \frac{\gm_1}{J+B} = \frac{\gm_2}{J+B} \; .
\ee
We consider the case of zero temperature and measure time in units
of $1/(J+B)$.

In this way, we come to the system of dimensionless equations describing
the tunneling intensity,
\be
\label{25}
\frac{dx}{dt} = ( 1 - b) yz - \gm ( x - x_t) \; ,
\ee
Josephson current,
\be
\label{26}
\frac{dy}{dt} = ( \om - x ) z - \gm y \; ,
\ee
and the well imbalance,
\be
\label{27}
\frac{dz}{dt} = (bx - \om ) y - \gm ( z - z_t) \; ,
\ee
with the local fields
\be
\label{28}
x_t = \frac{\om - bx}{h} \; , \qquad z_t = \frac{(1-b)z}{h} \; ,
\ee
where
\be
\label{29}
h = \sqrt{(\om - bx)^2 + (1-b)^2 z^2} \; .
\ee
Equations (25) to (29) form the main system of evolution equations
to be analyzed in what follows. We may notice that these equations
are invariant under the change of signs of $x$, $y$, and $\om$.
Fixing $\om$ positive implies that the energy levels in a double
well are enumerated so that $E_1<E_2$.

\section{Nonlinear dynamics}

First, let us consider the existence and stability of attractors of
Eqs. (25) to (27). We keep in mind that, by their definitions,
\be
\label{30}
\om \geq 0 \; , \qquad 0 \leq b < 1 \; .
\ee
There are two fixed points that are either stable or unstable depending
in the value of $\om$.

When the tunneling parameter is in the interval
\be
\label{31}
0 \leq \om < 1 \; ,
\ee
the stable fixed point is
\be
\label{32}
x_1^* = \om \; , \qquad y_1^* = 0 \; , \qquad
z_1^* =\sqrt{1-\om^2} \; .
\ee
The characteristic exponents, defined as the eigenvalues of the Jacobian
matrix, will be denoted by $\lbd$. For the fixed point (32), we have
\be
\label{33}
\lbd_1 = -\gm \qquad (\om < 1 ) \; ,
\ee
and two other exponents are the solutions of the quadratic equation
\be
\label{34}
(1 - b) \lbd^2 + \gm \left ( 2 - b -\om^2 \right ) \lbd +
\left ( 1 -\om^2 \right ) \left [ (1 - b)^2 + \gm^2
\right ] = 0 \; .
\ee
The general solutions of Eq. (34) are easily written, though they are
rather cumbersome. For illustration, we represent the characteristic
exponents for small $\gm <1$. Then
\be
\label{35}
\lbd_{2,3} \simeq -\; \frac{(2-b-\om^2)}{2(1-b)} \; \gm
\pm i \om_{eff} \; ,
\ee
with the effective frequency
\be
\label{36}
\om_{eff} = \sqrt{ ( 1 - b )\left ( 1 -\om^2\right )} \;
\left [ 1 \; - \; \frac{(b-\om^2)^2\gm^2}{8(1-b)^3(1-\om^2)}
\right ] \; .
\ee
This shows that the fixed point (32), for small $\gm$, is a stable focus.
The value $\om=1$ is a bifurcation point. When approaching this point,
the characteristic exponents $\lbd_2$ and $\lbd_3$ are
$$
\lbd_2 \simeq \left [ (1-b)^2 +\gm^2 \right ] \;
\frac{2(1-\om)}{(1-b)\gm} + \left [ ( 1 - b)^2 +\gm^2 \right ]
\left [ \gm^2 - 4 ( 1 - b)\right ] \; \frac{(1 -\om)^2}{(1-b)\gm^3} \; ,
$$
\be
\label{37}
\lbd_3 \simeq - \gm + 2 (1 - b) \frac{1-\om}{\gm} +
\left [ 4 (1 - b)^2 + ( 3 + b )\gm^2 \right ] \;
\frac{(1-\om)^2}{\gm^3} \; ,
\ee
for $\om\ra 1-0$. In the limit $\om=1$, the fixed point (32) is neutral,
since $\lbd_2=0$. When $\om$ grows larger than one, point (32) becomes
unstable.

For the tunneling parameter
\be
\label{38} \om > 1 \; ,
\ee
the stable fixed point is
\be
\label{39}
x_2^* = 1 \; , \qquad y_2^* = 0 \; , \qquad z_2^* =  0 \; .
\ee
For the characteristic exponents, we have
\be
\label{40}
\lbd_1 = - \gm \qquad (\om > 1)
\ee
and the equation
\be
\label{41}
\lbd^2 + \frac{2\om-1-b}{\om-b}\; \gm \lbd +
\frac{(\om-1)\left [ (\om-b)^2+\gm^2\right ]}{\om-b} = 0
\ee
defining $\lbd_2$ and $\lbd_3$. The latter for small $\gm$ are
\be
\label{42}
\lbd_{2,3} \simeq -\; \frac{2\om-1-b}{2(\om-b)}\; \gm \; \pm \;
i\om'_{eff} \; ,
\ee
with the effective frequency
\be
\label{43}
\om'_{eff} = \sqrt{(\om -1)(\om - b)} \;
\left [ 1 \; - \; \frac{(1-b)^2\gm^2}{8(\om-1)(\om-b)^3} \right ] \; .
\ee
Recall that $b<1$, hence $\om>b$. When $\om$ approaches the bifurcation
point from above, that is $\om\ra 1+0$, then
$$
\lbd_2 \simeq -\; \frac{(1-b)^2+\gm^2}{(1-b)\gm} \; (\om -1 ) \; -
\; \frac{\left[ (1-b)^2+\gm^2\right ]^2 -2\gm^4}{(1-b)^2\gm^3}\;
(\om -1)^2 \; ,
$$
\be
\label{44}
\lbd_3 \simeq - \gm + \frac{1-b}{\gm}\; (\om -1) +
\frac{(1-b)^2+2\gm^2}{\gm^3}\; (\om-1)^2 \; .
\ee
For $\om>1$, the fixed point (39) is stable. But at the bifurcation
point $\om=1$, the stationary solution (39) is neutral, since $\lbd_2=0$.
And point (39) becomes unstable for $\om<1$.

An important observation is that the system of equations (25), (26),
and (27) is structurally unstable if $\gm=0$, since then $\lbd_1=0$ and
$\lbd_2$ and $\lbd_3$ are purely imaginary. Such a behavior is rather
general for the cases not taking into account attenuation [31,32].
Allowance for the damping, due to particle collisions, makes the system
structurally stable and substantially influences the overall dynamics.
To emphasize the drastic change brought by the inclusion of the
attenuation, we solve Eqs. (25) to (27) numerically and present the
results in the figures.

Figure 1 shows the system dynamics for the tunneling parameter in the
region (31), when the stable fixed point is given by Eq. (32). The
initial conditions correspond to atoms loaded into the lattice in a
nonequilibrium state. If the attenuation is not taken into account,
the variables oscillate around their initial conditions. But giving
due consideration for the attenuation results in the variables relaxing
to the related stationary point.

In Fig. 2, the parameters correspond to the stable fixed point (39).
Again, at the initial time, atoms are loaded into the lattice in a
nonequilibrium state. When there is no damping, all variables oscillate
around their initial values. While, the presence of damping forces the
variables to relax to their stationary solutions.

Varying the parameter $b$ in the admissible region $0\leq b<1$ does
not influence much the system dynamics. Changing the tunneling parameter
$\om$ inside the region (31) also does not yield considerable difference
in the system behavior. When $\om>1$, then its increase results in more
pronounced oscillations of the dynamical variables. For comparison,
 Fig. 3 presents the system dynamics, with the same initial conditions
as in Fig. 2, but with a larger tunneling parameter.

\section{Conclusion}

An effective Hamiltonian for ultracold atoms in a double-well lattice
is derived. Nonequilibrium situation is considered, when atoms are
loaded into the lattice in a state far from a stationary one. As
dynamical variables, we analyze the tunneling intensity, Josephson
current, and the population displacement. The treatment is done in the
local-field approximation, taking into account the attenuation, caused
by particle collisions. It is shown that the proper account of damping
drastically changes the system dynamics, as compared to the case of no
attenuation included. If the simple mean-field approximation, with no
damping, is used, the dynamic variables oscillate around their initial
values. While taking into consideration the attenuation results in the
relaxation of the dynamical variables to their stationary solutions.
Depending on the value of the tunneling parameter, there exist two types
of stationary solutions, whose main difference is either in the presence
of a stationary atomic displacement in a double well, $z_1^*\neq 0$, or
in its absence, when the atoms in a double well are distributed
symmetrically, $z_2^*=0$.

An important message following from these results is that the so
popular simple mean-field approximation should be used with caution.
When considering nonlinear dynamics, such a simple approximation, with
no account of attenuation, can lead to qualitatively incorrect conclusions.

The results of the present paper are applicable to ultracold particles
of different nature. These could be neutral atoms and molecules [7--11]
or cold ions [33,34]. The basic requirement that the particles could be
loaded into a double-well lattice.

The double-well lattices possess a rich variety of properties, which
makes them an attractive candidate for different applications. A very
important point is that the lattice properties can be regulated and
controlled in a rather wide range. In addition to choosing the
appropriate values of the lattice parameters, it is also feasible to
realize their temporal modulation, similar to the case of usual optical
lattices, whose parameters can be made time dependent by applying
external fields [35--37], by moving [3,38], or just by shaking the
lattice [39,40].

\vskip 5mm

{\bf Acknowledgement}

\vskip 2mm

Financial support from the Russian Foundation for Basic Research (Grant
08-02-00118) is appreciated.

\newpage

\begin{center}
{\large{\bf Figure Captions}}

\end{center}

\vskip 5mm

{\bf Fig. 1}. Dimensionless variables describing the tunneling
intensity $x(t)$, Josephson current $y(t)$, and well population
imbalance $z(t)$ as functions of dimensionless time for $\om=0.1$
and $b=0.5$. Initial conditions are $x_0=0.66$, $y_0=0.75$, and
$z_0=0$. The case of no attenuation $(\gm=0)$ is shown by the dashed
curve. The case with attenuation $(\gm=1)$ is represented by the solid
line.

\vskip 5mm

{\bf Fig. 2}. Dimensionless variables $x(t)$, $y(t)$, and $z(t)$ as
functions of dimensionless time for $\om=1.5$ and $b=0.5$. Initial
conditions are $x_0=0.33$, $y_0=0.5$, and $z_0=0.8$. The attenuation
parameters are: $\gm=0$ (dashed line) and $\gm=1$ (solid line).

\vskip 5mm

{\bf Fig. 3}. Dimensionless tunneling intensity $x(t)$, Josephson
current $y(t)$, and population imbalance (population displacement)
$z(t)$ as functions of dimensionless time for $\om=5$ and $b=0.5$.
Initial conditions are $x_0=0.33$, $y_0=0.5$, and $z_0=0.8$. Behavior
is shown for two cases: no attenuation, $\gm=0$ (dashed line) and
with damping, $\gm=1$ (solid line).

\newpage

\begin{center}

\begin{figure}[hbtp]
\vspace{9pt}
\centerline{
\hbox{ \includegraphics[width=8cm]{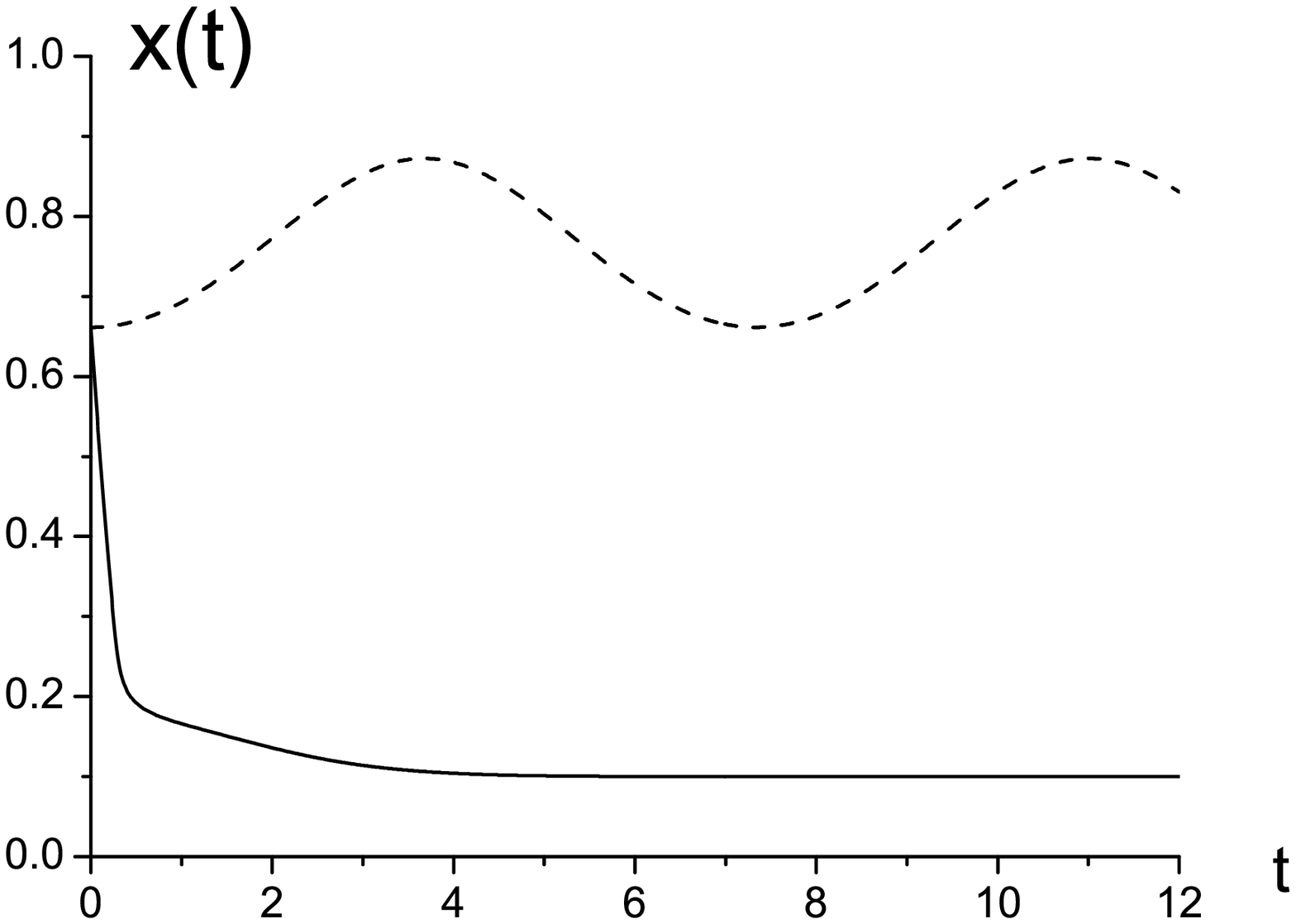} } }
\vspace{9pt}
\centerline{
\hbox{ \includegraphics[width=8cm]{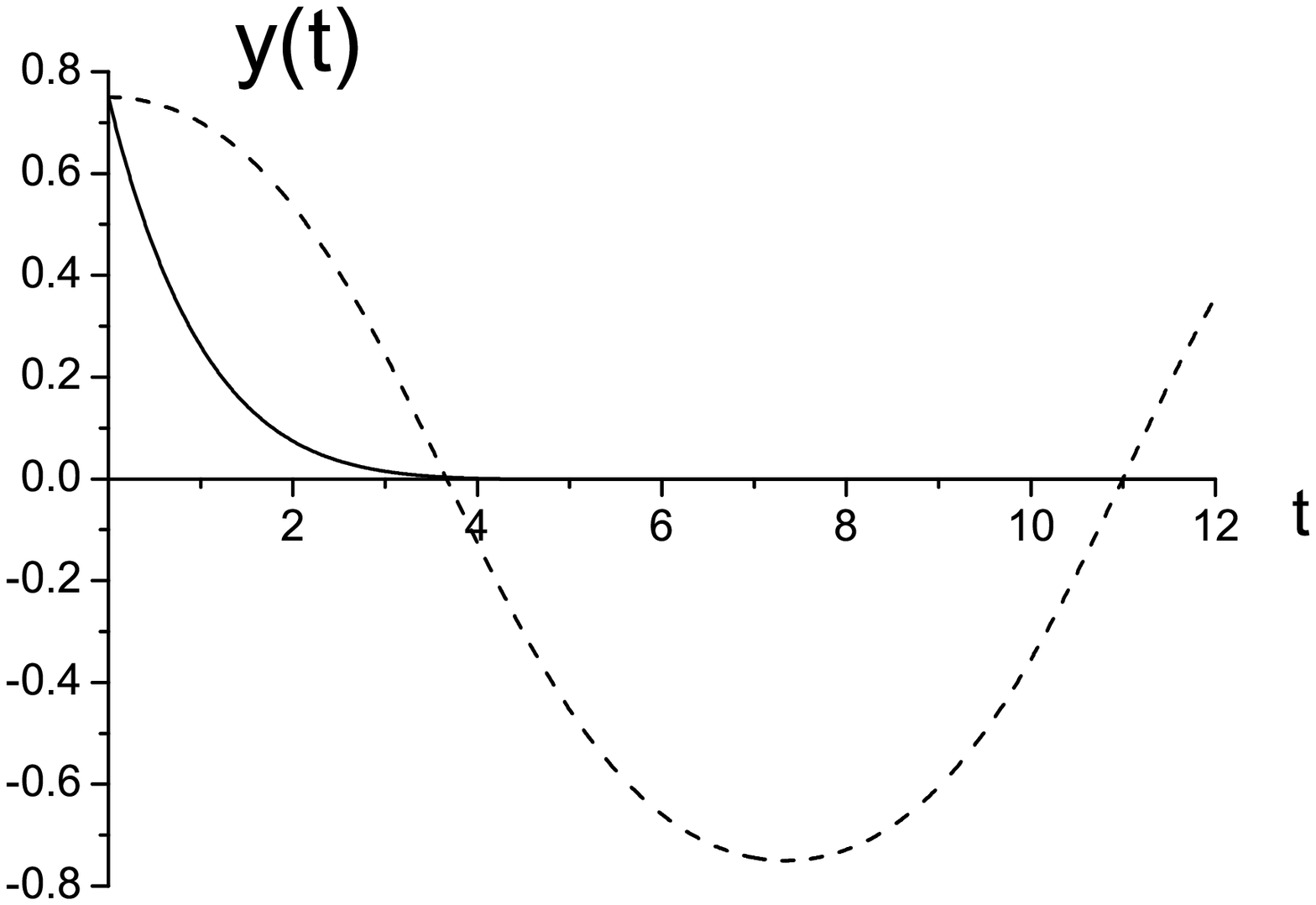} \hspace{1cm}
\includegraphics[width=8cm]{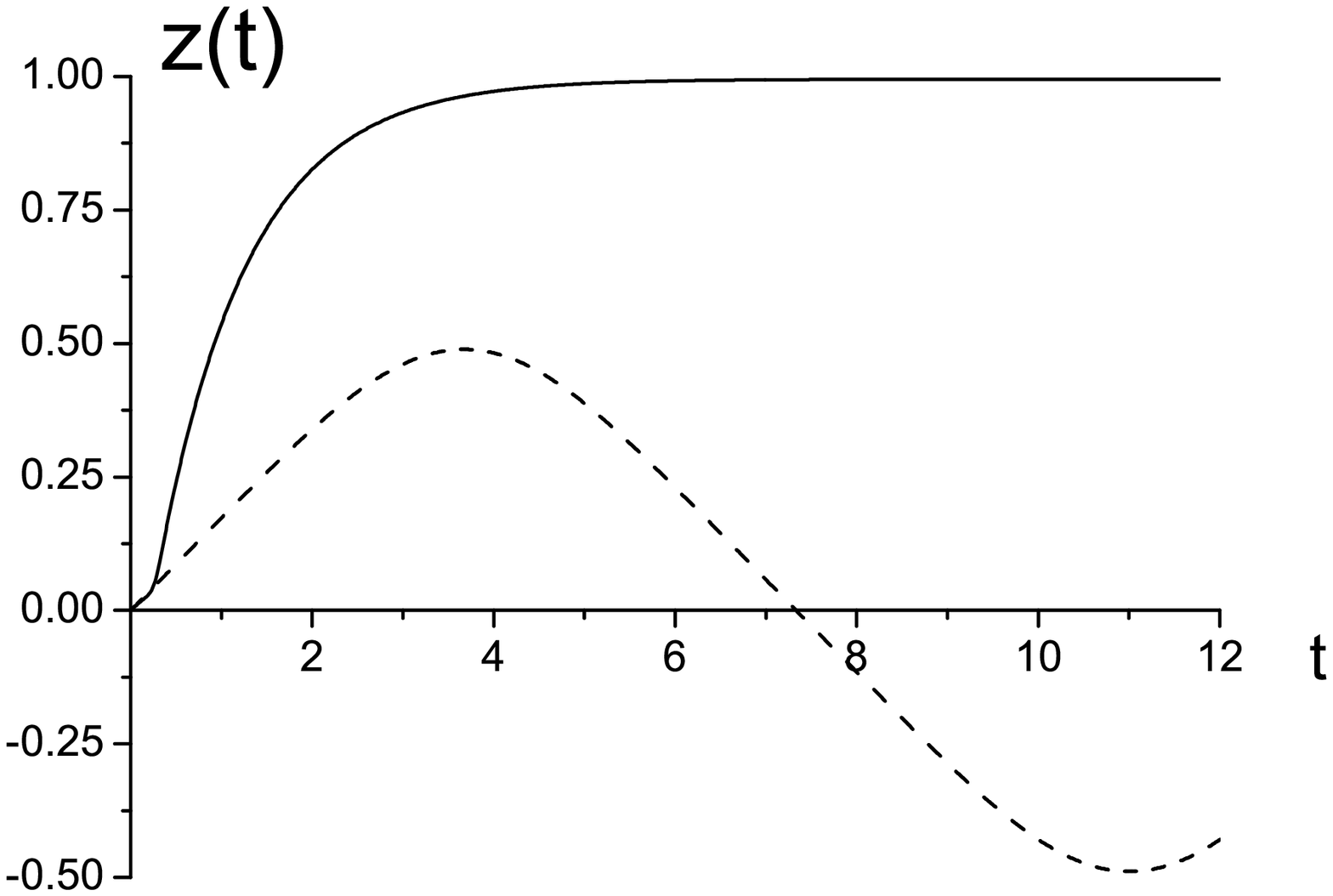} } }
\caption{Dimensionless variables describing the tunneling
intensity $x(t)$, Josephson current $y(t)$, and well population
imbalance $z(t)$ as functions of dimensionless time for $\om=0.1$
and $b=0.5$. Initial conditions are $x_0=0.66$, $y_0=0.75$, and
$z_0=0$. The case of no attenuation $(\gm=0)$ is shown by the dashed
curve. The case with attenuation $(\gm=1)$ is represented by the solid
line.}
\label{fig:Fig.1}
\end{figure}

\newpage

\begin{figure}[hbtp]
\vspace{9pt}
\centerline{
\hbox{ \includegraphics[width=8cm]{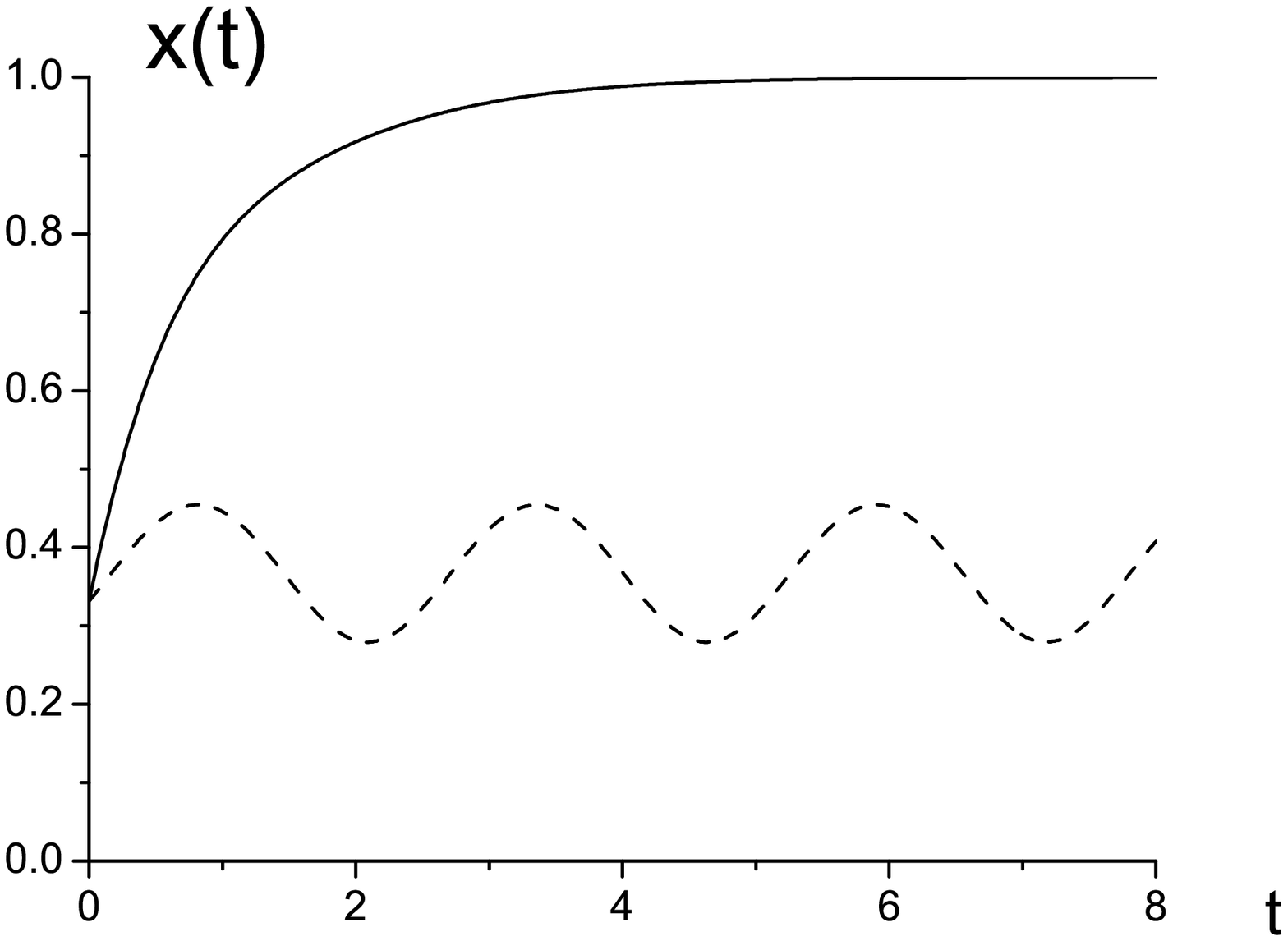} } }
\vspace{9pt}
\centerline{
\hbox{ \includegraphics[width=8cm]{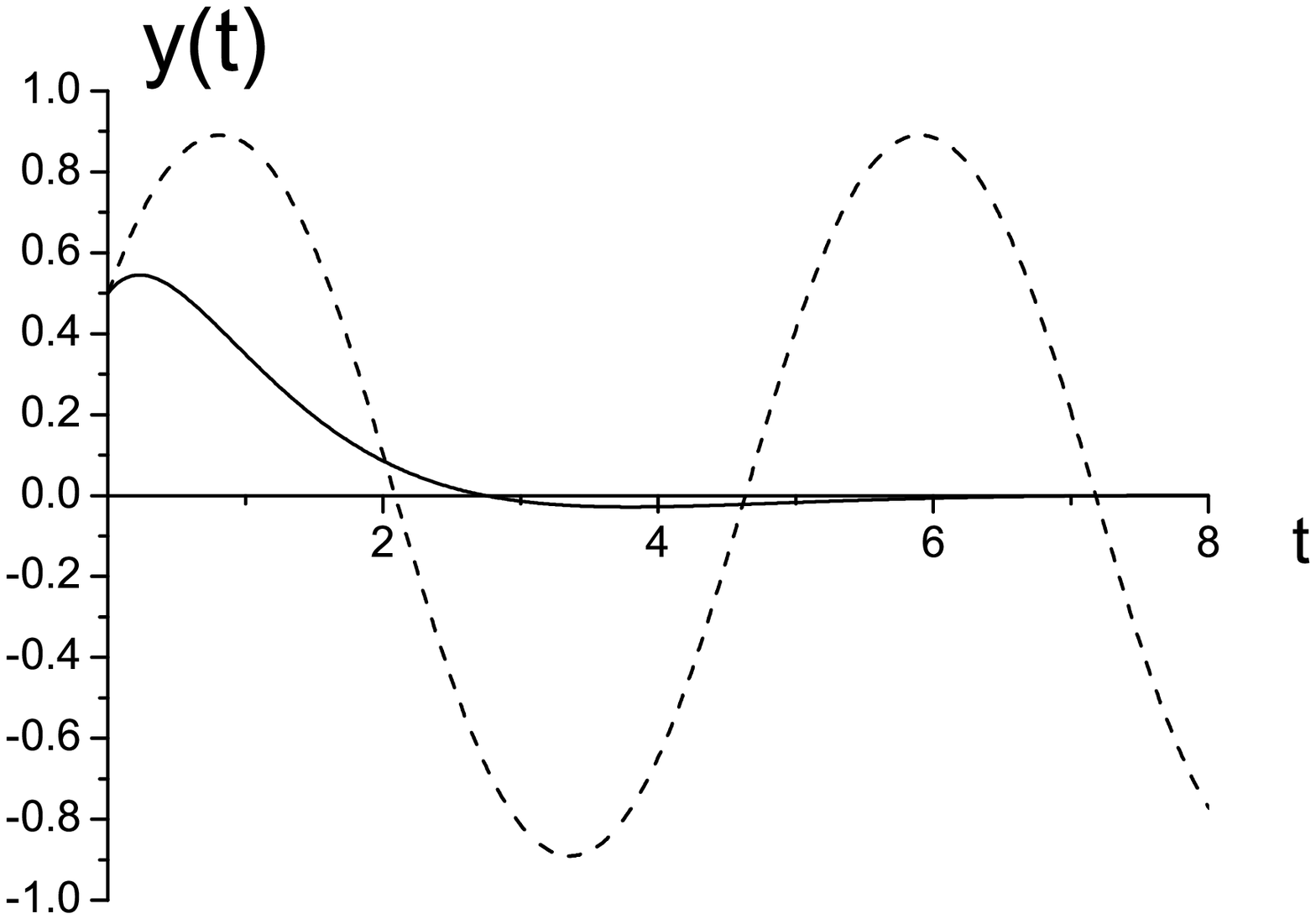} \hspace{1cm}
\includegraphics[width=8cm]{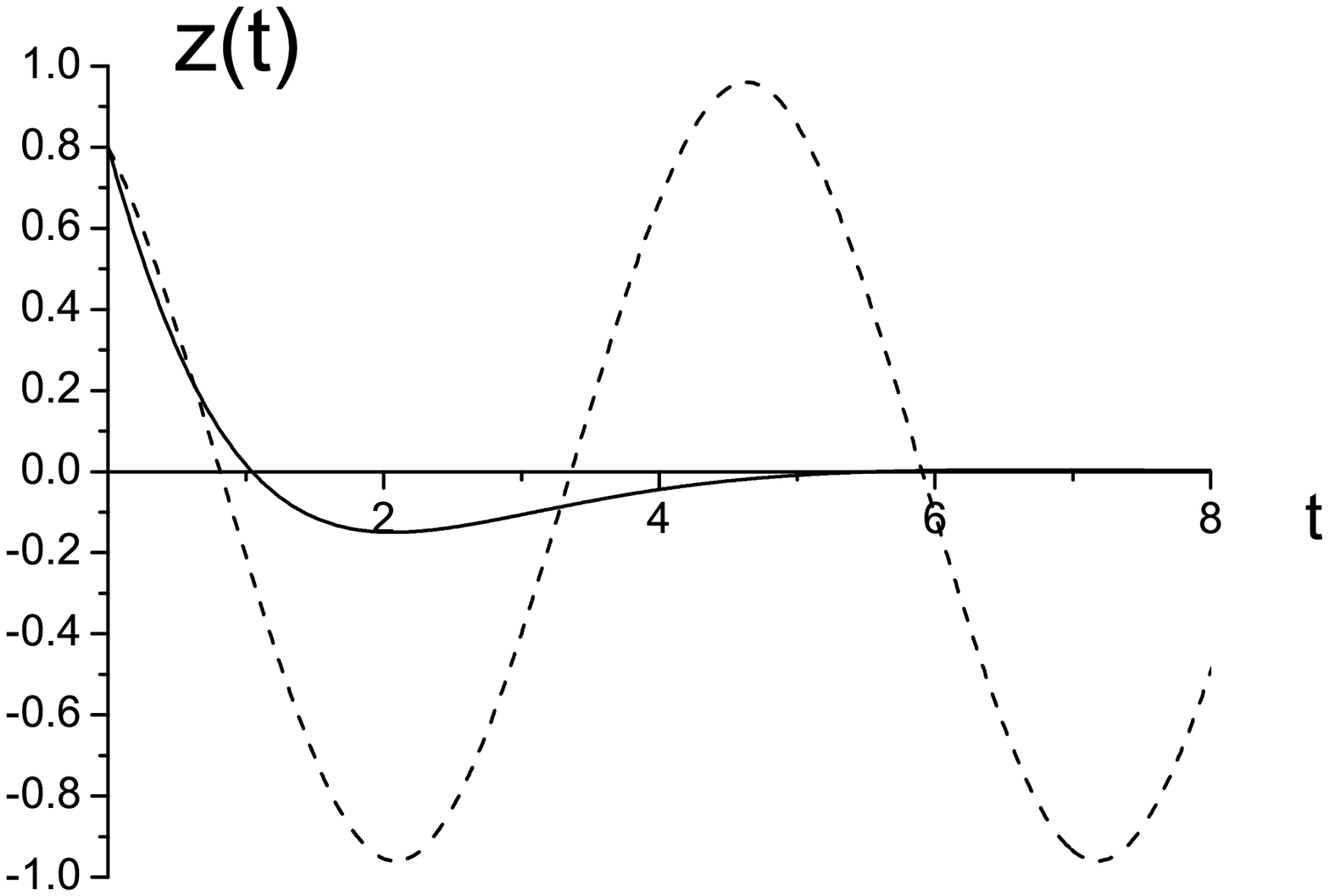} } }
\caption{Dimensionless variables $x(t)$, $y(t)$, and $z(t)$ as
functions of dimensionless time for $\om=1.5$ and $b=0.5$. Initial
conditions are $x_0=0.33$, $y_0=0.5$, and $z_0=0.8$. The attenuation
parameters are: $\gm=0$ (dashed line) and $\gm=1$ (solid line).}
\label{fig:Fig.2}
\end{figure}

\newpage

\begin{figure}[hbtp]
\vspace{9pt}
\centerline{
\hbox{ \includegraphics[width=8cm]{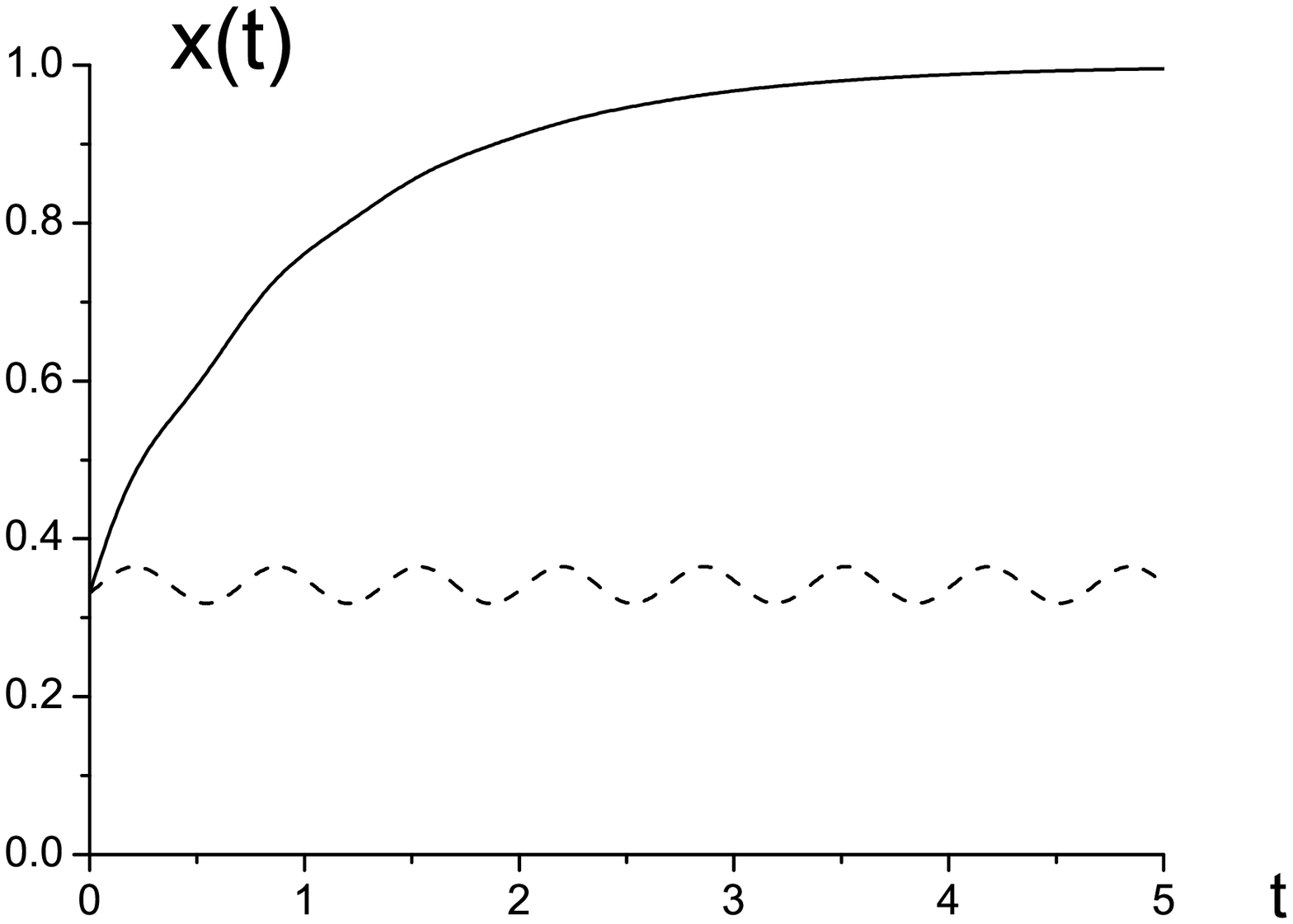} } }
\vspace{9pt}
\centerline{
\hbox{ \includegraphics[width=8cm]{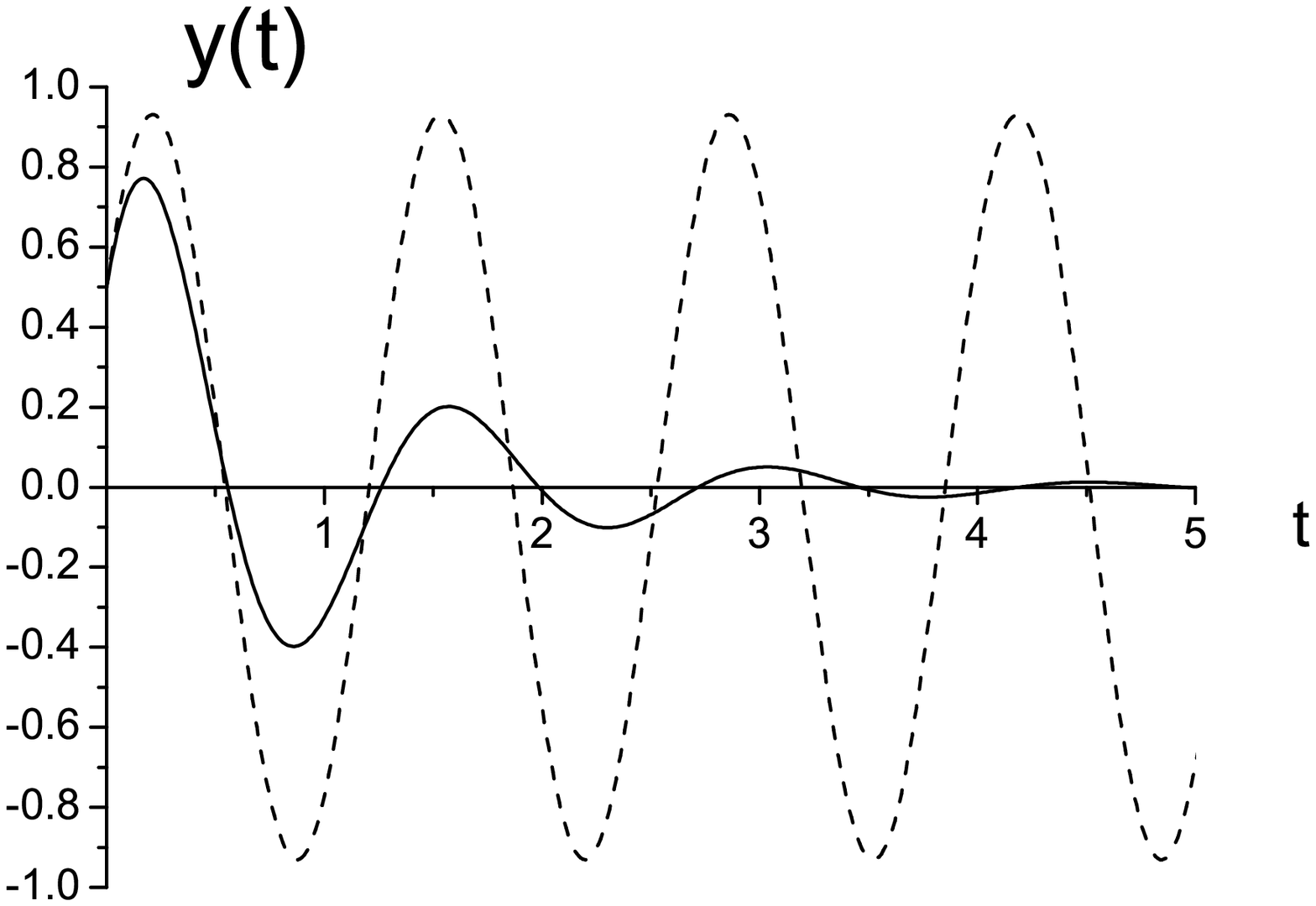} \hspace{1cm}
\includegraphics[width=8cm]{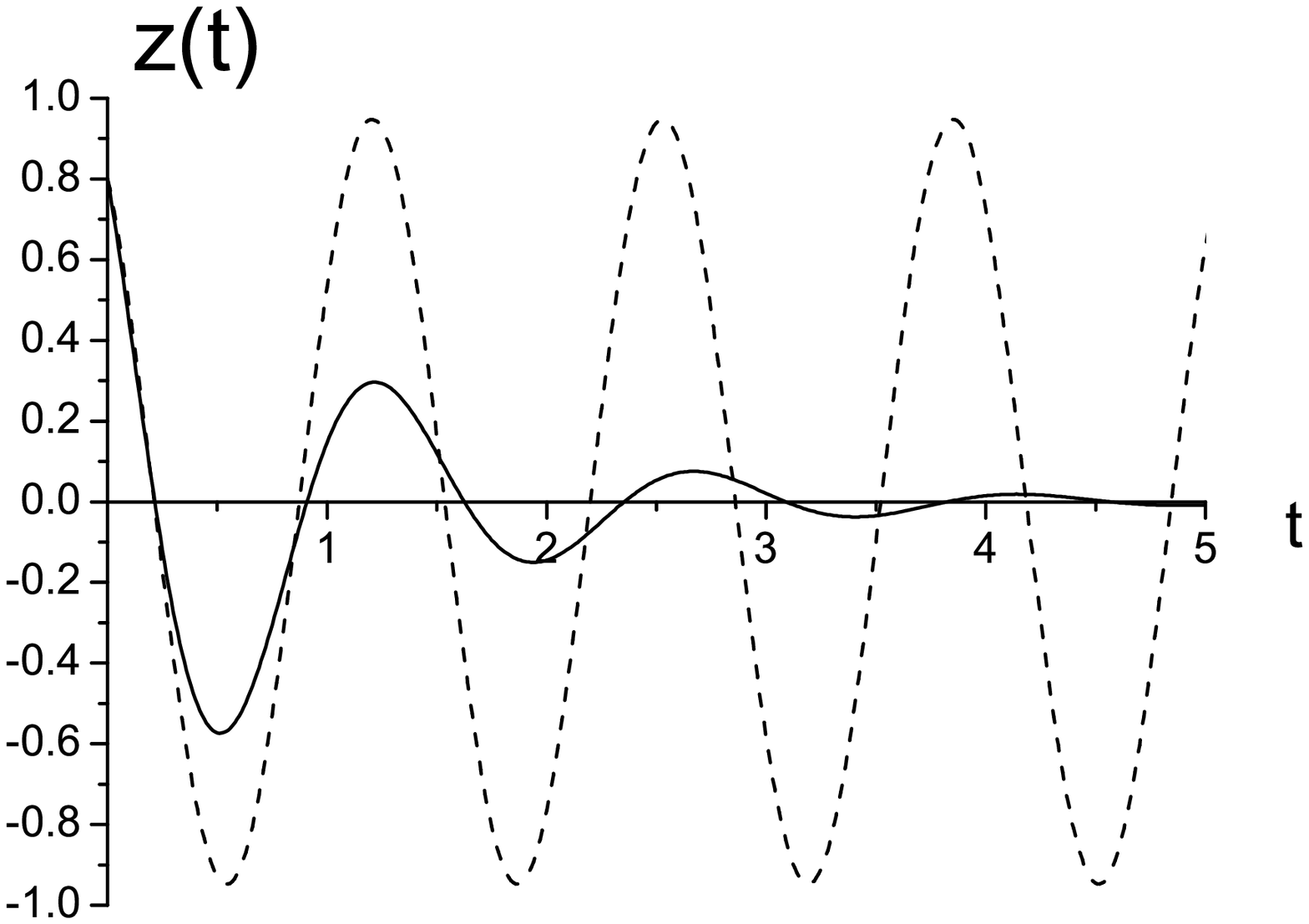} } }
\caption{Dimensionless tunneling intensity $x(t)$, Josephson
current $y(t)$, and population imbalance (population displacement)
$z(t)$ as functions of dimensionless time for $\om=5$ and $b=0.5$.
Initial conditions are $x_0=0.33$, $y_0=0.5$, and $z_0=0.8$. Behavior
is shown for two cases: no attenuation, $\gm=0$ (dashed line) and
with damping, $\gm=1$ (solid line).}
\label{fig:Fig.3}
\end{figure}

\end{center}

\end{document}